\documentclass[sn-mathphys,Numbered]{sn-jnl}
\usepackage{graphicx}%
\usepackage{multirow}%
\usepackage{amsmath,amssymb,amsfonts}%
\usepackage{amsthm}%
\usepackage{mathrsfs}%
\usepackage[title]{appendix}%
\usepackage{xcolor}%
\usepackage{textcomp}%
\usepackage{manyfoot}%
\usepackage{booktabs}%
\usepackage{algorithm}%
\usepackage{algorithmicx}%
\usepackage{algpseudocode}%
\usepackage{listings}%
\newtheorem{Theorem}{Theorem}
\newtheorem{Definition}{Definition}

\begin{document}

\title{Optimization of Quantum Measurements for Robustness Against Dark Counts: The $D$-Trace Optimality Framework}

\author*[1,2]{\fnm{Hao} \sur{Shu}}\email{Hao\_B\_Shu@163.com}

\affil*[1]{\orgname{Shenzhen University}}

\affil[2]{\orgname{South China University of Technology}}

\abstract{Quantum communication, while promising unparalleled security, faces significant practical challenges due to imperfections in quantum devices, particularly in single-photon detectors (SPDs). One of the key issues is the impact of dark counts, which cause erroneous detections and limit the effective range of quantum communication systems, such as in quantum key distribution (QKD). In this paper, we introduce a novel framework for addressing dark count errors through a new optimality criterion, termed $D$-trace optimality. This criterion aims to optimize quantum measurements to minimize the impact of dark counts without relying on hardware replacements. We propose an optimization scheme that transforms general measurements into $D$-trace optimal measurements, ensuring robustness against dark count effects. Our results suggest that, under certain conditions, this approach can improve the performance of quantum communication systems, especially in scenarios where the prior distribution of states is known. Although challenges remain in obtaining accurate prior distributions and ensuring compatibility with other strategies. This work lays the foundation for future research on optimizing quantum measurements to mitigate the effects of noise, including dark counts, in practice.}


\maketitle

\section{Introduction}
Quantum communication has emerged as a transformative technology with the potential to revolutionize secure communication protocols\cite{ZD2017Quantum,WL2022Quantum}, secret sharing\cite{HB1999Quantum,ZL2005Multiparty}, and cryptographic applications\cite{BB1984Quantum,LY2018Overcoming,S2022Measurement}. This field, rooted in the principles of quantum mechanics, has garnered significant attention over several decades, driven by its promise to achieve levels of security that are unattainable with classical communication systems. Despite the considerable theoretical advancements, the practical implementation of long-distance quantum communication remains fraught with challenges, primarily due to limitations inherent in real-world quantum devices\cite{QS2019Implementation,SL2021Deterministic}. One of the most pressing issues is the impact of dark counts in single-photon detectors (SPDs), which are crucial components of quantum communication systems\cite{H2009Single}, particularly in protocols like quantum key distribution (QKD).

Dark counts, which occur when a detector registers a signal despite no photon being present, introduce errors that severely limit the performance of long range quantum communication systems. These errors are particularly problematic in situations where the task requires classical outcomes, such as when the measured data is used to establish secure communication keys in QKD protocols\cite{BB2005Higher,SH2014Quantum}. As communication distances increase, the probability of photon loss and noise in the channel also rises, exacerbating the influence of dark counts. Thus, in practical implementations of quantum communication systems over lossy channels, including optical fibers\cite{WY2022Twin,TG2014Fundamental} and free-space channels\cite{AC2021Full}, the effective range of QKD systems is fundamentally constrained by dark counts in SPDs. In theory, if these dark counts could be eliminated or mitigated, the communication distance for repeaterless QKD could be extended infinitely\cite{S2022Solving}.

Despite the fundamental importance of addressing dark count errors, practical solutions have proven difficult to implement. The conventional approach to mitigating dark count errors often involves upgrading SPDs to more advanced technologies, such as superconducting detectors\cite{Y2020Superconducting, WM2009Superconducting, ZJ2019Saturating}, which are capable of achieving lower dark count rates compared to traditional ones\cite{JL20171.25,AW2004Efficient}. While superconducting detectors offer some promise, they come with significant drawbacks, including high cost, complex fabrication processes, and the need for operating conditions that are difficult to achieve and maintain $--$ such as extremely low temperatures (often below 1 Kelvin). Even with these sophisticated technologies, it is unlikely that dark counts can be completely eliminated. Moreover, the requirement for such advanced detectors makes them impractical for large-scale deployment, especially in field settings or for consumer-level applications. Consequently, there is an urgent need for alternative methods that can effectively mitigate the impact of dark counts without relying on the wholesale replacement of detection hardware.

In response to these challenges, the current paper proposes a novel and innovative approach to dealing with the dark count problem through the introduction of a new optimality criterion for quantum measurements, termed $D$-trace optimality. This criterion is specifically designed to address the impact of dark counts on quantum measurement outcomes, offering a new framework for optimizing the measurement strategies employed in quantum communication systems. The key innovation of this work lies in the introduction of an optimization scheme that identifies the most robust measurements against the effects of dark counts. Instead, the scheme focuses on transforming general quantum measurements into those that are optimal in terms of minimizing the influence of dark count errors.

The concept of $D$-trace optimality represents a significant departure from traditional methods of tackling dark count issues, which have typically focused on improving detector technology or reducing noise in the communication channel. By optimizing the measurement process itself, the proposed approach offers a potentially alternative solution in contrast to replacing or upgrading existing SPDs. To the best of our knowledge, this is the first attempt to address the dark count problem through the optimization of quantum measurements. Moreover, the proposed approach opens up new avenues for future research in the field of quantum communication. It invites further exploration into how different measurement strategies can be optimized to counteract specific sources of noise, including dark counts, in various quantum communication protocols. Additionally, it offers the potential to develop new methods for combining this optimization framework with other techniques, such as error correction or advanced signal processing, to further improve the performance of quantum communication systems.

\section{Motivation and Optimality}

This section introduces a framework to identify measurements that are robust against dark count effects, presenting a novel concept of optimality $--$ D-trace optimality.

\subsection{Motivation}

Consider a positive operator-valued measurement (POVM) $M=\{M_{k}|k=1,2,...,m\}$. In practical scenarios, the same POVM is often used repeatedly, making the statistical distribution of measurement outcomes a central concern. Suppose a set of quantum states $D$, each with prior probabilities, is measured. The average probability of observing an outcome corresponding to $M_{k}$ is given by $T_{k}=\int_{\rho\in D}Tr(M_{k}\rho)=Tr(M_{k}\int_{\rho\in D}\rho)$, where the integral averages over all states in $D$ with weighted. The term $\int_{\rho\in D}\rho$ represents the ensemble of $D$, namely describes a randomly chosen state in $D$, which is a mixed state and should be a constant related to $D$, in practice. A low $T_{k}$ indicates the corresponding effect is rarely used, reflecting its inefficiency. Note that strictly speaking, $T_{k}$ might depend on the measure of the space, however, this is not an important issue in this paper and thus assume that a suitable one is employed for simplicity.

The significance of $T_{k}$ becomes apparent in the presence of dark counts in single-photon detectors (SPDs). If the probability of detecting an outcome is much lower than the detector's dark count rate, the observed results will be heavily influenced by dark counts, leading to inaccuracies. Low-efficiency detectors, common in many practical settings, exacerbate this issue. Hence, maximizing $T_{k}$ minimizes the relative influence of dark count errors.

To illustrate, consider an SPD with dark count rate $\epsilon\prec 1$, measuring a state with an actual probability $T_{k}$ for effect $M_{k}$ in $m$ total effects. The observed probability is approximately $\frac{T_{k}+\epsilon}{1+m\epsilon} \approx T_{k}+\epsilon$, leading to a proportional of the observed probability and the actual probability approximating to $1+\frac{\epsilon}{T_{k}}$. To reduce the average error across all effects, one should minimize $\frac{\epsilon}{T_{k}}$, and on average in different effects, minimize $(\prod_{k=1}^{m}\frac{\epsilon}{T_{k}})^{\frac{1}{m}}$, which is achieved when all $T_{k}$ are equal.

\subsection{Definition}

Based on these considerations, $D$-trace optimality is formally defined:
\\

\begin{Definition}
	Let D be a set of states and $M=\{M_{k}|k=1,2,...,m\}$ be a not necessarily completed POVM (namely not necessarily satisfying $\sum_{k=1}^{m}M_{k}=I$), $M$ is said to be D-trace balance if $\int_{\rho\in D}Tr(M_{i}\rho)=\int_{\rho\in D}Tr(M_{j}\rho)$ for all i,j$\in \{1,2,...,m\}$, where the integral averages over all states in D. If in addition, $\sum_{k=1}^{m}Tr(M_{k}\rho)=1$ for all $\rho\in D$, then M is said to be D-trace optimal.
	
\end{Definition}
\vspace{0.5cm}
This definition ensures that the measurement outcomes are uniformly distributed over the state space $D$, minimizing susceptibility to dark count effects. Note that $\sum_{k=1}^{m}Tr(M_{k}\rho)=1$ represents that $M$ is completed when measuring states in $D$, namely if $M$ is completed by $M_{0}$, then the outcome $M_{0}$ is of null probability. An example of $\mathfrak{D}(C^{n})$-trace optimal POVM in $C^{n}$ is the SIC-POVM, where $\mathfrak{D}(C^{n})$ denotes the set of all density operators in $C^{n}$ with equal weight.

In the context of $C^{n}$, if $D=\mathfrak{D}(C^{n})$, the concept of $D$-trace optimality corresponds directly to the notion of unbiasedness, where unbiasedness for a single POVM means equal-trace\cite{DF2021The}. If all states are assigned equal weights, the trace of an effect accurately represents the probability of its occurrence. However, unbiasedness is an intrinsic property of a measurement that does not account for the distribution of the states being measured. By contrast, $D$-trace optimality explicitly incorporates the state distribution into the evaluation of optimality. In practical scenarios, the unknown state is often drawn from a known set with a specific distribution, rather than being uniformly distributed across the entire state space. As a result, unbiasedness alone is insufficient as a criterion, as the distribution of the states should influence the optimality of the measurement. For instance, while any rank-one projective measurement is unbiased, not all such measurements are necessarily optimal for a specific subset of states. Therefore, D-trace optimality can be regarded as a generalization of unbiasedness that takes into account prior information about the state distribution, enabling the design of more effective measurements in cases where the uniformity assumption does not hold.

\subsection{Non-uniform distribution}

In many practical scenarios, the state distribution is not uniform. For instance, in quantum state tomography, one might estimate parameters for an amplitude-damping channel\cite{YS2003Time} with Kraus operators:

$E_{0}=\begin{pmatrix}
	1 & 0 \\
	0 & \sqrt{1-p}
\end{pmatrix}$
and
$E_{1}=\begin{pmatrix}
	0 & \sqrt{p} \\
	0 & 0
\end{pmatrix}$, under the computational basis $\{|0\rangle, |1\rangle\}$. Here, $p$ represents the unknown noise strength, assumed to be low ($0\leq p\leq \delta$ for a small $\delta$). If the state $|1\rangle$ is transmitted through the channel, the resulting state becomes $p|0\rangle\langle 0|+(1-p)|1\rangle\langle 1|$. Estimating $p$ requires tomography on the non-uniform set $\{p|0\rangle\langle 0|+(1-p)|1\rangle\langle 1|\ |\ 0\leq p\leq \delta\}$. Such non-uniform distributions are common in real-world applications, highlighting the need for measurement optimization tailored to prior distributions.

\subsection{Optimalization scheme}

 The following scheme ensures any POVM can be transformed into a $D$-trace optimal POVM: 
 
 (1) Determine the maximum trace contribution: Compute $T=max_{k}(T_{k})$, for a given POVM $M=\{M_{1},...,M_{m}\}$.
 
 (2) Define the adjustment terms: Define $y_{k}=T-T_{k}\geq 0$. These terms measure the gap between the maximum trace $T$ and the contribution of each $M_{k}$ within $D$.
 
 (3) Construct the new POVM: Define the new POVM $M'=\{M_{k}'=\frac{M_{k}+y_{k}I}{mT}|k=1,2,...,m\}$ by the adjustment terms. Here, $I$ is the identity operator, and $m$ is the number of POVM elements. This step ensures the uniformity of the trace contributions over the subspace $D$, while maintaining the essential properties of the POVM.
 
 The theorem below establishes that $M'$ is a $D$-trace optimal POVM equivalent to $M$, where $'$equivalent$'$ means that the probabilities of $M_{k}'$ gives the probabilities of $M_{k}$ and conversely.
\\

\begin{Theorem}
	Every POVM M can be optimized to a D-trace optimal POVM M' using the above scheme.
\end{Theorem}
\vspace{0.5cm}
\noindent\textbf{Proof:}
\\

Let $M=\{M_{1},...,M_{m}\}$ be a POVM, and let $D$ represent a subspace of quantum states. Define $T_{k}=\int_{\rho\in D}Tr(M_{k}\rho)$, $T=max_{k}(T_{k})$, and $y_{k}=T-T_{k}\geq 0$. The new POVM is defined as $M'$ satisfies the properties of a $D$-trace optimal POVM and is equivalent to $M$.
\\

\textbf{Equivalence of $M$ and $M'$:}

The equivalence between $M$ and $M'$ arises from the uniform rescaling and the addition of the adjustment term $y_{k}$. The probability of obtaining any measurement outcome $M_{k}$ when applied to a quantum state is preserved in $M'$, as shown below, noting that $y_{k}$ is fixed.

$\int_{\rho\in D}Tr(M_{k}'\rho)=\int_{\rho\in D}Tr(\frac{M_{k}+y_{k}I}{mT}\rho)=Tr(\frac{\int_{\rho\in D} M_{k}\rho+y_{k}\int_{\rho\in D} \rho}{mT})=\frac{1}{mT}(T_{k}+y_{k})$.
\\

\textbf{Positivity of $M'$:}

Each $M_{k}'=\frac{M_{k}+y_{k}I}{mT}$ is positive semi-definite because $M_{k}$ as well as $y_{k}I$ is positive semi-definite (with $y_{k}\geq 0$), while the normalization factor $mT\geq 0$ does not affect the positivity.
\\

\textbf{Uniform trace contribution over $D$:}

For any $k$, $\int_{\rho\in D}Tr(M_{k}'\rho)=\frac{1}{mT}(T_{k}+y_{k})=\frac{1}{mT}(T_{k}+T-T_{k})=\frac{T}{mT}=\frac{1}{m}$.
\\

\textbf{Normalization of $M'$:}

Summing over all $M_{k}'$, we have:
$\sum_{k=1}^{m}M_{k}'=\sum_{k=1}^{m}\frac{M_{k}+y_{k}I}{mT}=\frac{\sum_{k=1}^{m}M_{k}+(\sum_{k=1}^{m}y_{k})I}{mT}$.

By the definition of a POVM, $\sum_{k=1}^{m}M_{k}=I$ and,

$\sum_{k=1}^{m}T_{k}=\sum_{k=1}^{m}\int_{\rho\in D}Tr(M_{k}\rho)=\int_{\rho\in D}Tr((\sum_{k=1}^{m}M_{k})\rho)=\int_{\rho\in D}Tr(\rho)=1$. 

Thus, $\sum_{k=1}^{m}y_{k}=\sum_{k=1}^{m}(T-T_{k})=mT-\sum_{k=1}^{m}T_{k}=mT-1$. 

Therefore, $\sum_{k=1}^{m}M_{k}'=\frac{\sum_{k=1}^{m}M_{k}+(\sum_{k=1}^{m}y_{k})I}{mT}=\frac{I+(mT-1)I}{mT}=I$ and,

 $\sum_{k=1}^{m}Tr(M_{k}'\rho)=Tr(\sum_{k=1}^{m}M_{k}')\rho)=Tr(\rho)=I$.
\\

We conclude that the constructed POVM $M'$ satisfies all the conditions of a $D$-trace optimal POVM and the theorem has been proven.

$\hfill\blacksquare$
\vspace{0.5cm}
\section{Discussion}

\subsection{Additional justifications}

A potential concern might arise regarding the direct subtraction of the dark count rate from each probability of effects to mitigate the impact of dark counts in a POVM. In practice, however, the dark count rate of a detector is typically known only within a certain range, and it may vary over time with use. Consequently, accurately subtracting the dark count rate would necessitate frequent estimation, which could increase operational costs and shorten the detector's lifespan, even if such estimation is feasible.

 Alternatively, using a fixed estimated value within the range for subtraction may not be fully effective. For instance, suppose the actual dark count rate of a detector is $\epsilon\prec 1$ but the estimated dark count rate used for subtraction is $\eta$, while the true probability of an effect remains $T_{k}$. In this case, the ratio of the observed probability to the actual probability approximates $1+\frac{\epsilon-\eta}{T_{k}}$. It is important to note that the magnitude of $\epsilon-\eta$ is generally of the same order as $\epsilon$, meaning that $\epsilon-\eta$ is typically as small as $\epsilon$. For example, if the prior knowledge suggests that $0\leq\epsilon\leq 10^{-4}$, an appropriate estimation might be $\eta=0.5\times 10^{-4}$.

 Another intuitive justification for trace-optimality, although not strictly formal, is that every device has a limited operational lifespan. If we assume, for simplicity, that each detector corresponds to a single effect and has a fixed, uniform lifetime (though real-world configurations may be more complex), then the number of states that can be detected is constrained by the most frequently employed effect. For example, consider a device that can be used 100 times before it ceases to function. If the device employs a POVM $M=\{M_{1}, M_{2}, M_{3}, M_{4}\}$ with equal effect probabilities $T_{1}=T_{2}=T_{3}=T_{4}=0.25$, the device could measure 400 states on average. However, if the device employs a POVM $M=\{M_{1}, M_{2}, M_{3}, M_{4}\}$ with probabilities $T_{1}=0.05,\ T_{2}=0.05,\ T_{3}=0.1,\ T_{4}=0.8$, can only measure 125 states on average. the device could only measure 125 states on average. In this case, the efficiency of the device, defined as the maximum number of states it can measure on average, is limited by the largest value of $T_{k}$. Thus, to optimize efficiency, it is desirable to minimize the largest $T_{k}$ and, ideally, to equalize all $T_{k}$'s.
 
 \subsection{Limitations and future works}
 The $D$-trace optimality framework allows for the incorporation of the prior distribution of states to effectively mitigate dark count issues in quantum measurements. By utilizing prior knowledge of the state distribution, this approach aims to optimize the measurement process, thereby reducing the impact of dark counts on the overall performance of the system. However, it is important to note that the framework may not offer advantages in all contexts. Specifically, the $D$-trace optimality criterion may not be the most effective approach for scenarios where the prior distribution of states is either unknown or difficult to estimate accurately. In such cases, alternative optimization strategies might be more appropriate. Therefore, the $D$-trace framework may be particularly beneficial when employed alongside measurements that have already been optimized from other perspectives. For example, optimization schemes based on frame potential\cite{BF2003Finite} or uncertainty volume minimization\cite{WFOptimal1989} are valuable in situations where the goal is to achieve a balance between measurement consumption and robustness. These complementary strategies can be combined with the $D$-trace optimality framework to enhance the overall performance of quantum measurements, especially in systems where the prior distribution is known or can be reasonably estimated.
 
 On the other hand, while the $D$-trace approach may be particularly effective when the prior distribution of states is available, obtaining such a distribution in real-world applications remains a significant challenge. In practice, prior distributions are often difficult to determine, especially in dynamic or complex systems where the states being measured may vary over time or be subject to unknown fluctuations. This limitation can hinder the full applicability of the $D$-trace optimality framework, especially in cases where precise knowledge of the state distribution is crucial. As a result, future research may need to focus on developing methods to estimate or approximate the prior distribution, thereby expanding the range of applications for the $D$-trace approach.

Another key limitation of the optimization scheme presented prior to Theorem 1 is that the effects in the optimized POVM $M'$ differ from those in the original POVM $M$, which could introduce difficulties in certain tasks. In particular, this discrepancy may affect scenarios where the projective nature of the original POVM $M$ is a critical requirement. For example, $M$ may consist of projective effects that ensure precise state discrimination or preserve specific mathematical or physical properties, whereas $M'$, as a result of optimization, does not necessarily retain these projective effects. This deviation can introduce uncertainty in the precision of measurement outcomes and potentially compromise the interpretability of results in contexts where projective measurements are expected or necessary. Moreover, the lack of projective structure in $M'$ could limit its applicability to tasks that inherently rely on projection-based outcomes, such as quantum error correction, quantum state preparation, or specific quantum information protocols. In these cases, the optimized POVM $M$ might not be directly compatible with the theoretical or experimental frameworks that assume projectivity, thereby requiring additional modifications or recalibration.

In practical implementations, the difference between $M$ and $M'$ could also result in added complexity when designing measurement apparatuses or aligning experimental setups to match the optimized POVM. Such adjustments may increase operational costs or introduce new sources of error, potentially offsetting the benefits of optimizing for $D$-trace optimality. As a result, while the optimization scheme is effective in reducing the influence of dark counts, its implications for broader applicability and compatibility with existing projective schemes should be carefully considered.

\section{Conclusion}

In this work, we have proposed a novel optimization scheme based on the concept of $D$-trace optimality, specifically designed to mitigate the impact of dark counts in quantum communication systems. By optimizing the measurement process, this framework offers a promising alternative to hardware-dependent solutions. The core advantage of the D-trace approach lies in its ability to incorporate the prior distribution of states, allowing for more robust quantum measurements that are less susceptible to dark count errors. This is particularly valuable for long-distance quantum communication, where noise and photon loss in the channel exacerbate dark count issues.

However, while the framework shows significant promise, there are limitations that must be addressed. In particular, obtaining accurate prior distributions of states remains a challenge in real-world applications, especially in dynamic systems where the state distribution may vary over time. Furthermore, the optimization scheme may not be universally advantageous in all contexts, particularly in cases where the prior distribution is unknown or difficult to estimate.

Future work should focus on improving methods for estimating prior distributions, developing adaptive optimization techniques, and exploring how $D$-trace optimality can be integrated and implemented with other strategies, such as error correction. Moreover, the potential of $D$-trace optimality must be carefully considered in tasks where precise measurement outcomes are critical, as the optimized measurements may not retain the projectivity. 

Despiteness, the D-trace optimality framework represents a promising step forward in addressing the dark count problem and enhancing the performance of quantum communication systems in practical settings.

\section{Statements and Declarations}

\subsection{Data Availability}

No Data associated in the manuscript.

\subsection{Competing Interests}

The author declares no competing interests.

\subsection{Author contributions}

The paper has a unique author.

  \bibliography{Bibliog}

	\end{document}